\documentclass [12pt]{article}
\usepackage[a4paper, margin=0.9in]{geometry}

\usepackage{graphicx}		
\usepackage{amsmath}				
\usepackage{bm}						
\usepackage{float}
\usepackage{epstopdf}				
\usepackage{natbib}					
\usepackage{amssymb}				
\usepackage{longtable}				
\usepackage{verbatim}				
\usepackage{kbordermatrix}
\usepackage{url}
\usepackage{lscape}
\usepackage{xcolor}

\usepackage[margin=20pt, font=small, labelfont=bf]{caption}

\title{}
\author{}

\begin{document}

\title{The 2015-2017 policy changes to the means-tests of Australian Age Pension: implication to decisions in retirement}
\author{Johan G. Andr\'{e}asson \footnote{CSIRO, Australia; School of Mathematical and Physical Sciences, University of Technology, Sydney, Broadway, PO Box 123, NSW 2007, Australia; email: johan.andreasson@uts.edu.au} , Pavel V. Shevchenko \footnote{Applied Finance \& Actuarial Studies, Macquarie University; email: pavel.shevchenko@mq.edu.au}}
\date{\today}
\maketitle

\begin{abstract}
The Australian Government uses the means-test as a way of managing the pension budget. Changes in Age Pension policy impose difficulties in retirement modelling due to policy risk, but any major changes tend to be `grandfathered' meaning that current retirees are exempt from the new changes. In 2015, two important changes were made in regards to allocated pension accounts -- the income means-test is now based on deemed income rather than account withdrawals, and the income-test deduction no longer applies. 
We examine the implications of the new changes in regards to optimal decisions for consumption, investment, and housing. We account for regulatory minimum withdrawal rules that are imposed by regulations on allocated pension accounts, as well as the 2017 asset-test rebalancing. The new policy changes are modelled in a utility maximizing lifecycle model and solved as an optimal stochastic control problem. We find that the new rules decrease the benefits from planning the consumption in relation to the means-test, while the housing allocation increases slightly in order to receive additional Age Pension. The difference in optimal drawdown between the old and new policy are only noticeable early in retirement until regulatory minimum withdrawal rates are enforced. However, the amount of extra Age Pension received for many households is now significantly different due to the new deeming income rules, which benefit slightly wealthier households who previously would receive no Age Pension due to the income-test and minimum withdrawals. \\


\end{abstract}

\newpage

\section{Introduction}
Australia relies on a defined-contribution pension system that is based on the superannuation guarantee, private savings, and a government provided Age Pension. The superannuation guarantee mandates that employers contribute a set percentage of the employee's gross earnings to a superannuation fund, which accumulates and is invested until retirement. The current contribution rate is set to 9.5\%, where contributions in addition to this often comes with tax benefits. Private savings comprise of these contributions, but also include savings outside the superannuation fund such as investment accounts, dwelling, and other assets. Finally, the Age Pension is a government managed safety net which provides the retiree with a means-tested Age Pension. This means-test determines whether the retiree qualifies for full, partial, or no Age Pension once the entitlement age is reached. In the means-test, income and assets are evaluated individually, and a certain taper rate reduces the maximum payments once income or assets surpass set thresholds (which are subject to family status and homeownership). Income from different sources are also treated differently; financial assets are expected to generate income based on a progressive deeming rate, while income streams such as labor and non account-based annuity payments are assessed based on their nominal value.

Since the Australian retirement system is relatively young, the long-term effects of this new pension system are not yet known. Changes in this system are expected to occur frequently due to fiscal reasons, and once the effects policy changes have on a retiree's personal wealth (and the economy in general) becomes evident. Variables directly related to the means-test such as entitlement age, means-test thresholds, taper rates, and pension payments can all be adjusted to meet budget needs by the government. On a larger scale, regulatory changes may include whether the family home is included in the means-tested assets, the elimination of minimum withdrawal\footnote{Certain account types for retirement savings have a minimum withdrawal rate once the owner is retired.} rules, changes in mandatory savings rates, or additional taxes on superannuation savings. From a mathematical modelling perspective, this poses difficulties in terms of future model validity, as regulatory risk and policy changes can quickly make a model obsolete if it is not modified to account for the new rules. 

The motivation for this paper was the recent changes for allocated pension accounts, where assets now generate a deemed income and which no longer have an income-test deduction. Account-based pensions (such as allocated pension accounts) are accounts that have been purchased with superannuation and generate an income stream throughout retirement. Prior to 2015, these types of accounts allowed for an income-test deduction that was determined upon account opening, and withdrawals were considered to be income in the means-test. The income-test deduction allowed the retiree to withdraw slightly more every year without missing out on Age Pension. However, in 2015 the rules changed. Existing accounts were `grandfathered' and will continue to be assessed under the old rules, while the new rules will be applied to any new accounts. The argument for the changes were simplicity (people with the same level of assets should be treated the same no matter how they are invested), to increase incentive to maximize total disposable income rather than maximizing Age Pension payments, and to level how capital growth and interest paying investments were assessed \citep{Fahcsia2016}. From a fiscal point of view, the recommendations to introduce the new rules were based on estimated unchanged costs\footnote{The recommendations to introduce deeming was made in \cite{Henry2009} where the fiscal sustainability is evaluated with the general equilibrium model `KPMG Econtech MM900' \citep{KPMG2010}. The model shows the estimation over a 10-year window hence we do not know the short term or year-to-year estimates. In addition to this, the model includes additional suggested tax and budget related changes, hence the effect of introducing deeming rates cannot be isolated.} \citep{Henry2009}, however the 2015-2016 budget stated expected savings of \$57m for 2015-2016, and \$129m and \$136m for subsequent years \citep{TheCommonwealthofAustralia2015}. The Age Pension post in the 2015-2016 budget includes all changes to the Age Pension in a combined viewpoint, so a specific impact of the deeming rule changes is not known. We adapt the model previously developed in \cite{Andreasson2016} to examine the impact of this policy change on an individual retiree. The model used is an extension with stochastic factors (mortality, risky investments and sequential family status), to what was originally presented in \cite{Ding2013, Ding2014}, which is an expected utility model for the retirement behavior in the decumulation phase of Australian retirees subject to consumption, housing, investment, bequest and government provided means-tested Age Pension.

Problems with decisions that span over multiple time periods are typically modelled with lifecycle models and solved with backward recursion (\citeauthor{Cocco2005}, \citeyear{Cocco2005}; \citeauthor{Cocco2012}, \citeyear{Cocco2012}; \citeauthor{Blake2014}, \citeyear{Blake2014} to name a few). While there is a plethora of research on the subject internationally, there is still rather limited research modelling the Australian Age Pension, and even less that enforces the minimum withdrawal rules. The original model in \cite{Ding2013} does not constrain drawdown with minimum withdrawal, which would limit the author from finding a closed form solution. Similarly, other authors that focus on means-tested pension also do not enforce minimum withdrawal rates, such as \cite{Hulley2013} who use CRRA utility to understand consumption and investment behavior, or \cite{Iskhakov2015} who investigate how annuity purchases changes in relation to Age Pension. It should be noted that their assumptions do not include Allocated Pension accounts, thus minimum withdrawal rates may not apply. There is surprisingly limited research conducted on implications of the regulatory minimum withdrawal rates, even though a large number of retirees are using such accounts (or similar phased withdrawal products). The exception is \cite{Bateman2007}, who compare the welfare of retirees when the current minimum withdrawal rates were introduced 2007 against the previous rules and alternative drawdown strategies. The authors use a rather simple CRRA model to examine the effect of different risk aversion and investment strategies but find that the minimum withdrawal rules increase the welfare for retirees although slightly less than optimal drawdown does. In \cite{Andreasson2016} the minimum withdrawal rules are included in part of the model outcome, but is by no means exhaustive and only provides a brief introduction to the effects. 

The minimum withdrawal rules are designed to exhaust the retiree's account around year 100, however after year 85 (subject to investment returns) the withdrawn dollar amount starts decreasing quickly. In a recent report from \cite{PFL2016} it is identified that only 5\% of retirees exhaust their accounts completely, though this number is expected to increase as life expectancy increases and the population ages. They find that retirees tend to follow the minimum withdrawal rules as guidelines for their own withdrawal, as few withdraw more than the minimum amount. This is further confirmed in \cite{Shevchenko2016}. Even so, \cite{RiceWarner2015} argues that the minimum withdrawal rates should be cut by 25-50\% to prevent retirees from exhausting their superannuation prematurely due to increased longevity. The current rates are simply too high for many retirees, thus is not sustainable for people living longer than the average life expectancy,\, and are significantly higher than what is optimal in \cite{Andreasson2016}.

The contribution of this paper is to improve the understanding of the effect the new policy rules, and with minimum withdrawal enforced, has on a typical retiree's optimal decisions. Both policy rules are implemented in a utility maximization model with stochastic mortality and risky investments, as well as sequential family status, which explains the behavior of Australian retirees well. We then examine the differences in optimal decisions between an allocated pension account opened prior to 2015 with the one opened post 2015, as well as compare with the planned 2017 asset-test adjustments and the previous results where minimum withdrawal is not enforced. The paper is structured as follows: In Section 2 we summarize the model and present the Age Pension function, as well as explain the parameterization. Section 3 contains a discussion of the results. Finally, in Section 4 we present our concluding remarks. 

\section{Model}
The model utilized is from \cite{Andreasson2016}, where the Age Pension function has been updated to account for the policy changes in 2015. For a complete description of the model, its calibration to the data and numerical solution, and a discussion of the construction and assumptions, please see the reference. 

The objective of the retiree is to maximize expected utility generated from consumption, housing, and bequest. The retiree starts off with a total wealth $\mathsf{W}$, and at the time of retirement $t=t_0$ is given the option to allocate wealth into housing $H$ (and if he already is a homeowner, the option to adjust current allocation by up- or downsizing). The remaining (liquid) wealth $W_{t_0} = \mathsf{W} - H$ is placed in an allocated pension account, which is a type of account that does not have a tax on investment earnings and is subject to the regulatory minimum withdrawal rates. A retiree can either start as a couple or single household, where this information is contained in a family state variable

\begin{equation}
G_t \in \mathcal{G} = \{\Delta, 0, 1, 2\},
\end{equation}
where $\Delta$ corresponds to the agent already deceased at time $t$, $0$ corresponds to the agent died during $(t-1,t]$, $1$ and $2$ correspond to the agent being alive at time $t$ in a single or couple households respectively. Evolution in time of the family state variable $G_t$ is subject to survival probabilities. In the case of a couple household, there is a risk each time period that one of the spouses passes away, in which case it is treated as a single household model for the remaining years.

At the start of each year $t=t_0, t_0+1...,T-1$ the retiree will receive a means-tested Age Pension $P_t$, and decide what amount of saved liquid wealth $W_t$ will be used for consumption (defined as proportion drawdown $\alpha_t$ of liquid wealth), and the proportion $\delta_t$ of remaining liquid wealth that will be invested in risky assets. The change in wealth after the decision to next period is then defined as
\begin{equation}
\label{eq:transition}
W_{t+1} = \left[ W_t - \alpha_t W_t \right] \left[\delta_t e^{Z_{t+1}} + (1-\delta_t) e^{r_t} \right],
\end{equation}
where $Z_{t+1}$ is the stochastic return on risky assets modelled as independent and identically distributed random variables from Normal distribution $\mathcal{N}(\mu-\widetilde{r},\sigma)$ with mean $\mu - \widetilde{r}$, variance $\sigma^2$ and inflation\footnote{By defining the model in real terms (adjusted for inflation), time-dependent variables do not have to include inflation which otherwise would be an additional stochastic variable.} rate $\widetilde{r}$. Any wealth not allocated to risky assets is assumed to generate a deterministic real risk-free return $r_t$ (risk-free interest rate adjusted for inflation).
Each period the agent receives utility based on the current state of family status $G_t$:
\begin{equation}
\label{RewardFunction}
 R_{t}(W_t,G_t,\alpha_t,H) = \left\{ \begin{array}{ll}
         U_C(C_t,G_t,t) + U_H(H,G_t), & \mbox{if $G_t = 1,2$},\\
         U_B(W_t), & \mbox{if $G_t = 0$},\\
         0, & \mbox{if $G_t = \Delta.$}\end{array} \right. 
\end{equation} 
That is if the agent is alive he receives reward (utility) based on consumption $U_C$ and housing $U_H$, if he died during the year the reward comes from the bequest $U_B$, and if he is dead there is no reward. Note that the reward received when the agent is alive depends on whether the family state is a couple or single household due to differing utility parameters and Age Pension thresholds. 

Finally, $t=T$ is the maximum age of the agent beyond which survival is deemed impossible, and the terminal reward function is given as

\begin{equation}
\label{TerminalRewardFunction}
\widetilde{R}(W_T,G_T) = \left\{ \begin{array}{ll}
         U_B(W_T), & \mbox{if $G_T \geq 0,$}\\
         0, & \mbox{if $G_T = \Delta.$}\end{array} \right. 
\end{equation} 

The retiree has to find the decisions that maximize expected utility with respect to the decisions for consumption, investment, and housing. This is defined as a stochastic control problem, where decisions (controls) at time $t$ depend on stochastic variable realization at time $t$ but where future realizations are unknown. The problem can be defined as

\begin{equation}
\label{eq:FinalValueFunction}
\underset{H}{\max} \left[ \underset{\boldsymbol{\alpha}, \boldsymbol{\delta}}{\sup} \: \mathbb{E}^{\boldsymbol{\alpha}, \boldsymbol{\delta}}_{t_0} \left[\beta_{t_0,T} \widetilde{R}(W_T,G_T) + \sum_{t={t_0}}^{T-1} \beta_{t_0,t} R_{t}(W_t,G_t,\alpha_t,H)\right] \right],
\end{equation}
 where $\mathbb{E}^{\alpha, \delta}_{t_0}[\cdot]$ is the expectation conditional on information at time $t=t_0$ if we use control $\boldsymbol{\alpha} = (\alpha_{t_0}, \alpha_{t_0+1}, ..., \alpha_{T-1})$ and $\boldsymbol{\delta} = (\delta_{t_0}, \delta_{t_0+1}, ..., \delta_{T-1})$ for $t=t_0, t_0+1, ..., T-1$. The subjective discount rate $\beta_{t,t'}$ is a proxy for personal impatience between time $t$ and $t'$. This problem can be solved numerically with dynamic programming by using backward induction of the Bellman equation. The state variables are discretized on a grid, and the Gaussian Quadrature method is used for integration between periods; for details, see \cite{Andreasson2016}.

\subsection{Utility functions}
Utility in the model is measured with time-separable additive functions based on the commonly used HARA utility function, subject to different utility parameters for singles and couples, as follows.

\begin{itemize}
\item{} \textbf{Consumption preferences}. It is assumed that utility comes from consumption exceeding the consumption floor, weighted with a time-dependent health status proxy\footnote{Note that the purpose is not to model health among the retirees, but rather to explain decreasing consumption with age.}. The utility function for consumption is defined as

\begin{equation}
\label{eq:consumption}
U_C (C_t,G_t,t) = \frac{1}{\psi^{t-t_0} \gamma_d} \left(\frac{C_t - \overline{c}_d}{\zeta_d} \right)^{\gamma_d}, \quad d = \left\{ \begin{array}{ll}
         \mathrm{C}, & \mbox{if $G_t = 2 \quad \text{(couple),}$}\\
         \mathrm{S}, & \mbox{if $G_t = 1 \quad \text{(single),}$}\end{array} \right.
\end{equation}
where $\gamma_d \in (-\infty,0)$ is the risk aversion and $\overline{c}_d$ is the consumption floor parameters. The scaling factor $\zeta_d$ normalizes the utility a couple receives in relation to a single household. The utility parameters $\gamma_d$, $\overline{c}_d$ and $\zeta_d$ are subject to family state $G_t$, hence will have different values for couple and single households. Also, $\psi \in [1,\infty)$ is the utility parameter for the health status proxy, which controls the declining consumption between current time $t$ and time of retirement $t_0$. 

\item \textbf{Bequest preferences}. Utility is also received from \emph{luxury} bequest, hence the home is not included in the bequest \citep{Ding2014}. The utility function for bequest is then defined as

\begin{equation}
U_B(W_t) = \left(\frac{\theta}{1-\theta}\right)^{1-\gamma_\mathrm{S}} \frac{\left(\frac{\theta}{1-\theta} a+W_t\right)^{\gamma_\mathrm{S}}}{\gamma_\mathrm{S}},
\end{equation}
where $W_t$ is the liquid assets available for bequest, and $\gamma_\mathrm{S}$ the risk aversion parameters for single households\footnote{The risk aversion is considered to be the same as consumption risk aversion for singles since a couple is expected to become a single household before bequeathing assets.}. The parameter $\theta \in \left[0,1\right)$ is the degree of altruism which controls the preference of bequest over consumption, and $a \in \mathbb{R}^+$ is the threshold for luxury bequest up to where the retiree leaves no bequest\footnote{Because the marginal utility is constant for the bequest utility with zero wealth, in a model with perfect certainty and CRRA utility the optimal solution will suggest consumption up to level $a$ before it is optimal to save wealth for bequest \citep{Lockwood2014}.}.

\item \textbf{Housing preferences}. The utility from owning a home comes in the form of preferences over renting but is approximated by the home value. The housing utility is defined as
\begin{equation}
U_H (H,G_t) = \frac{1}{\gamma_\mathrm{H}} \left(\frac{\lambda_d H}{\zeta_d} \right)^{\gamma_\mathrm{H}},
\end{equation}
where $\gamma_\mathrm{H}$ is the risk aversion parameter for housing (allowed to be different from risk aversion for consumption and bequest), $\zeta_d$ is the same scaling factor as in equation (\ref{eq:consumption}), $H > 0$ is the market value of the family home at time of purchase $t_0$ and $\lambda_d \in (0,1]$ is the preference of housing defined as a proportion of the market value. 
\end{itemize}

\subsection{Age Pension}
The Age Pension policy changes over time, and all income streams of  allocated pension accounts opened after the 1st January  2015 are assumed to generate deemed income. Accounts opened prior to this are `grandfathered' hence will continue to be assessed under the old rules \citep{Fahcsia2016}, where instead drawdown is considered income.

The Age Pension rules state that the entitlement age is 65 for both males and females, with the means-test thresholds and taper rates for July 2016 presented in Table \ref{table:PensionRates} and discussed in detail later in this section. 
The new rules have introduced a `Work bonus' deduction for the income-test, but as the model assumes the retiree is no longer in the workforce this has been left out.

\begin{table}[!h]
\caption{Age Pension rates published by Centrelink as at September 2016.}
\label{table:PensionRates}
\centering
\begin{tabular}{l l c c}
\hline
\multicolumn{3}{r}{Single} & Couple\\
\hline
$P^d_\mathrm{max}$ & Full Age Pension per annum & \$22,721 & \$34,252\\
\hline
& \bf{Income-Test}\\
$L^{d}_\mathrm{I}$ & Threshold & \$4,264 & \$7,592\\
$\varpi^d_{\mathrm{I}}$ & Rate of Reduction & \$0.5 & \$0.5\\
\hline
& \bf{Asset-Test}\\
$L^{d,h=1}_\mathrm{I}$ & Threshold: Homeowners & \$209,000 & \$296,500\\
$L^{d,h=0}_\mathrm{I}$ & Threshold: Non-homeowners & \$360,500 & \$448,000\\
$\varpi^d_{\mathrm{A}}$ & Rate of Reduction & \$0.039 & \$0.039\\
\hline
& \bf{Deeming Income}\\
$\kappa^d$ & Deeming Threshold & \$49,200 & \$81,600\\
$\varsigma_-$ & Deeming Rate below $\kappa^d$& 1.75\% & 1.75\%\\
$\varsigma_+$ & Deeming Rate above $\kappa^d$& 3.25\% & 3.25\%\\
\hline
\end{tabular}
\end{table}

\subsubsection{Deemed income}
Deemed income refers to the assumed returns from financial assets, without reference to the actual returns on the assets held. The deemed income only applies to financial assets and account based income streams and is calculated as a progressive rate of assets. The income-test can therefore depend on both labor income (if any), deemed income from financial investments not held in the allocated pension account, drawdown from allocated pension accounts if opened prior to 2015, or deemed income on such accounts if opened after January 1\textsuperscript{st} 2015.

The deeming rates are subject to change in relation to interest rates and stock market performance\footnote{The current rates are at a historical low. In 2008 the deeming rates $\varsigma_{-} /\varsigma_{+}$ were as high as 4\%/6\%, but in March 2013 they were set to 2.5\%/4\% due to decreasing interest rates, then in November 2013 to 2\%/3.5\% and to current levels of 1.75\%/3.25\% in March 2015.}. Two different deeming rates may apply based on the value of the account; a lower rate $\varsigma_{-}$ for assets under the deeming threshold $\kappa_d$ and a higher rate $\varsigma_{+}$ for assets exceeding the threshold, as shown in Table \ref{table:PensionRates}.

\subsubsection{Age Pension function}
The Age Pension received is modelled with respect to the current liquid assets, where the account value is used for the asset-test. Since the model assumption states that no labor income is possible, all income for the income-test comes from either deemed income (new rules) or generated from withdrawals of liquid assets (old rules). The Age Pension function can thus be defined as

\begin{equation}
P_t := f(W_t) = \max \left[0, \min \left[P^{d}_\mathrm{max}, \min \left[P_{\mathrm{A}}, P_{\mathrm{I}}\right] \right] \right],
\end{equation}
where $P^d_{\mathrm{max}}$ is the full Age Pension, $P_\mathrm{A}$ is the asset-test and $P_\mathrm{I}$ is the income-test functions. The $P_\mathrm{A}$ function is the same for rules prior and post 2015, and is defined as

\begin{equation}
P_{\mathrm{A}}:= P^d_\mathrm{max} - (W_t-L^{d,h}_{\mathrm{A}})\varpi^d_{\mathrm{A}},
\end{equation}
where $L_\mathrm{A}^{d,h}$ is the threshold for the asset-test and $\varpi^d_\mathrm{A}$ the taper rate for assets exceeding the thresholds. Superscript $d$ is a categorical index indicating couple or single household status as defined in equation (\ref{eq:consumption}). The variables are subject to whether it is a single or couple household, and the threshold for the asset-test is also subject to whether the household is a homeowner or not ($h=\{0,1\}$).
Although the $P_{\mathrm{A}}$ function is the same for both the old and new policies, the $P_{\mathrm{I}}$ function is different. For the new policy rules, it can be written as

\begin{equation}
P_{\mathrm{I}}:= P^d_\mathrm{max} - (P_{\mathrm{D}}(W_t) - L^{d}_{\mathrm{I}})\varpi^d_{\mathrm{I}},
\end{equation}
\begin{equation}
P_{\mathrm{D}}(W_t) = \varsigma_- \min \left[ W_t, \kappa^d \right] + \varsigma_+ \max \left[0, W_t - \kappa^d \right],
\end{equation}
where $L_\mathrm{i}^d$ is the threshold for the income-test and $\varpi^d_\mathrm{I}$ the taper rate for income exceeding the threshold. $P_{\mathrm{D}}(W_t)$ calculates the deemed income, where $\kappa_d$ is the deeming threshold, and $\varsigma_{-}$ and $\varsigma_{+}$ are the deeming rates that apply to assets below and above the deeming threshold respectively. 

Under the previous policy, the $P_I$ function is defined as

\begin{equation}
P_{\mathrm{I}}:= P^d_\mathrm{max} - (\alpha_t W_t - M(t) -  L^{d}_{\mathrm{I}})\varpi^d_{\mathrm{I}},
\end{equation}
\begin{equation}
M(t) = \frac{W_{t_0}}{e_{t_0}}(1+\widetilde{r})^{t_0-t},
\end{equation}
where the function $M(t)$ represents the income-test deduction that was available for accounts opened prior to 2015, $e_{t_0}$ is the lifetime expected at age $t_0$ and $\widetilde{r}$ the inflation. As the model is defined in real terms, the future income-test deductions must discount inflation. Current values of the function parameters are given in Table \ref{table:PensionRates}. 

\subsection{Parameters}
The model parameters are taken from \cite{Andreasson2016}, where calibration was performed on empirical data from \cite{Statistics2011}. However, the consumption floor $\overline{c}_d$ and the threshold for luxury bequest $a$ must be adjusted as they represent monetary values. Since the previous model was defined in real terms, we need to set a new base year for the comparison. We therefore adjust these parameters based on the Age Pension adjustments from 2010 to 2016. Currently, the Age Pension payments are adjusted to the higher of the Consumer Price Index (CPI) and Male Average Weekly Total Earnings (MTAWE). The increase in full Age Pension payments from 2010 to 2016 equals approximately 4.5\% increase per year. We assume that the utility parameters representing monetary values have increased in the same manner. All utility model parameter values are shown in Table \ref{table:parameters}.

\begin{table}[h]
\centering
\caption{Model parameter values adjusted for 2016}
\label{table:parameters}
\begin{tabular}{l c c c c c c c c c}
\hline
& $\gamma_d$ & $\gamma_\mathrm{H}$ & $\theta$ & $a$ & $\overline{c}_d$ & $\psi$ & $\lambda$ & $\zeta_d$\\
\hline
Single household & -1.98 & -1.87 & 0.96 & \$27,200 & \$13,284 & 1.18 & 0.044 & 1.0\\
Couples household & -1.78 & -1.87 & 0.96 & \$27,200 & \$20,607 & 1.18 & 0.044 & 1.3\\
\hline
\end{tabular}
\end{table}

On the 1\textsuperscript{st} of January 2017 the thresholds of the asset-test will be `rebalanced', hence will change significantly \citep{AustralianGovernmentDepartmentofVeteransAffairs2016}. The thresholds for the asset-test-will be increased and the taper rate $\varpi^d_{\mathrm{A}}$ will double. This effectively means that retirees will receive full Age Pension for a higher level of wealth, but once the asset-test binds, the partial Age Pension will decrease twice as fast, causing them to receive no Age Pension at a lower level of wealth than before. At the time of writing of this paper there are no proposed adjustments to the full Age Pension or income-test threshold for January 2017, hence these Age Pension parameters do not have to be adjusted other than updating the asset-test thresholds and taper rate according to the changes. The parameters for the Age Pension in 2016 are shown in Table \ref{table:PensionRates}, and the 2017 Age Pension parameters for the updated asset-test are shown in Table \ref{table:PensionRates2017}. In addition to this, we set the following.
\begin{itemize}
\item A retiree is eligible for Age Pension at age $t=65$ and lives no longer than  $T=100$. 
\item Real risky returns follow $Z_t \sim \mathcal{N}(0.056, 0.133)$, and the real risk-free rate is set to $r_t = 0.005$. These parameters were estimated from S\&P/ASX 200 Total Return and the deposit rate \citep{Andreasson2016}.
\item The lower threshold for housing is set to \$30,000. A retiree with wealth below this level can therefore not be a homeowner, hence $H\in \{0,[30000, \mathrm{W}]\}$.
\item A unisex survival probability is used to avoid separating the sexes. The survival probabilities for a couple are assumed to be mutually exclusive, based on the oldest partner in the couple. The actual mortality probabilities are taken from Life Tables published in \cite{ABSMortality2014}.
\item The subjective discount rate $\beta$ is set in relation to the real interest rate so that $\beta_{t,t'} = e^{-\sum_{i=t}^{t'}r_i}$.
\end{itemize}

\begin{table}[!h]
\caption{Planned 2017 Age Pension rates published by Centrelink as at September 2016 (\emph{https://www.humanservices.gov.au/customer/services/centrelink/age-pension}).}
\label{table:PensionRates2017}
\centering
\begin{tabular}{l l c c}
\hline
\multicolumn{3}{r}{Single} & Couple\\
\hline
& \bf{Asset-Test}\\
$L^{d,h=1}_\mathrm{I}$ & Threshold: Homeowners & \$250,000 & \$375,000\\
$L^{d,h=0}_\mathrm{I}$ & Threshold: Non-homeowners & \$450,000 & \$575,000\\
$\varpi^d_{\mathrm{A}}$ & Rate of Reduction & \$0.078 & \$0.078\\
\hline
\end{tabular}
\end{table}

Minimum withdrawal rates for allocated pension accounts are shown in Table \ref{table:minwithdrawal} \citep{ATO2016}. The rates impose a lower bound on optimal consumption, hence withdrawals from liquid wealth must be larger or equal to these rates.

\begin{table}[!h]
\centering
\caption{Minimum regulatory withdrawal rates for allocated pension accounts for the year 2016 and onwards (\emph{https://www.ato.gov.au/rates/key-superannuation-rates-and-thresholds/}).}
\label{table:minwithdrawal}
\begin{tabular}{l c c c c c c c}
\hline
Age & $\le$ 64 & 65-74 & 75-79 & 80-84 & 85-89 & 90-94 & 95 $\le$ \\
Min. drawdown & 4\% & 5\% & 6\% & 7\% & 9\% & 11\% & 14\% \\
\hline
\end{tabular}
\end{table}

\section{Results}
In the income-test, the policy change to replace asset drawdown with deemed income leads to some interesting implications for the retirees in all three decision variables (housing, consumption, and risky asset allocation). The main difference is that assets are now included twice in the means-test, as the income-test is now based on assets only rather than the actual drawdown of assets. Optimal decisions are then becoming more sensitive to changes in liquid assets, although the retiree has now less control to optimize utility in relation to the Age Pension.

Below we present and compare results for optimal decisions under the old rules (`Pre 2015'), new deeming rules for income-test (`Post 2015'), and new deeming rules with the new asset test (`Post 2015, new asset test') that starts in 2017. 

\subsection{Optimal Consumption}
\begin{figure}[!h]
  \centering
  \includegraphics[width=.9\linewidth]{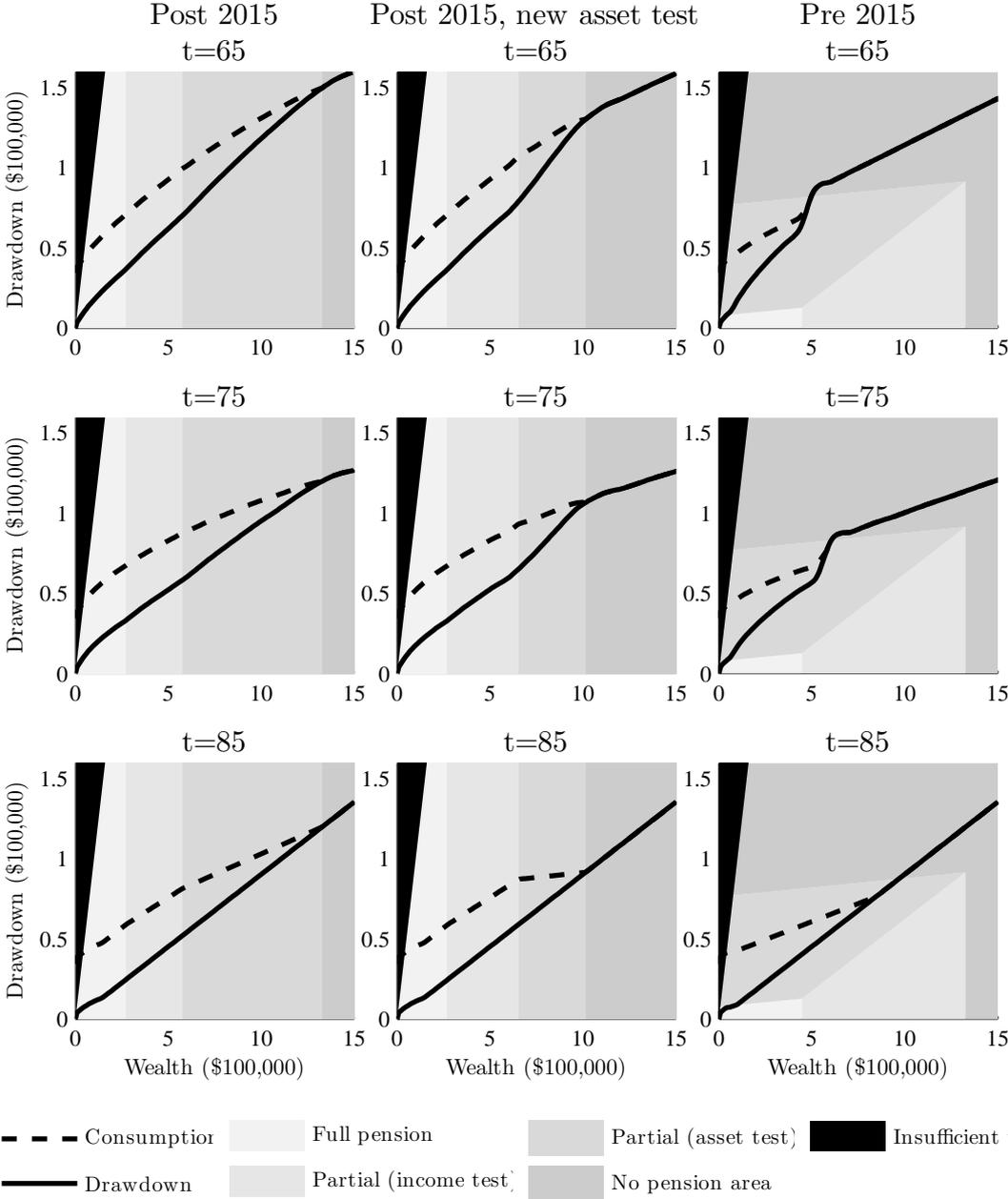}
  \captionof{figure}{Optimal drawdown and consumption for non-homeowner couple households under the three different policy scenarios.}
  \label{fig:OptDD_C_W}
\end{figure}

The optimal consumption consists of the drawdown from liquid wealth and the Age Pension received, and exemplifies a behavior consistent with traditional utility models (Figure \ref{fig:OptDD_C_W}). The curve is generally a smooth, concave, and monotone function of wealth hence becomes flatter as wealth increases due to decreasing marginal utility. The curve becomes flatter as the retiree ages, which is the desired effect from the model's health proxy as to reflect the lower consumption resulting from decreasing health. However, this general behavior starts to deviate as the retiree ages due to the minimum withdrawal rates. For a retiree aged 65 with an account of \$500,000, the optimal consumption for a couple is roughly 11.1\% which is more than the minimum withdrawal rate of 5\% (Table \ref{table:minwithdrawal}). As the retiree ages his consumption tends to decrease, but around age 85 the minimum withdrawal rates crosses over the optimal consumption hence the drawdown curve becomes proportional to wealth. This deviation occurs at an even earlier age for wealthier retirees.

\begin{figure}[!h]
  \centering
  \includegraphics[width=.9\linewidth]{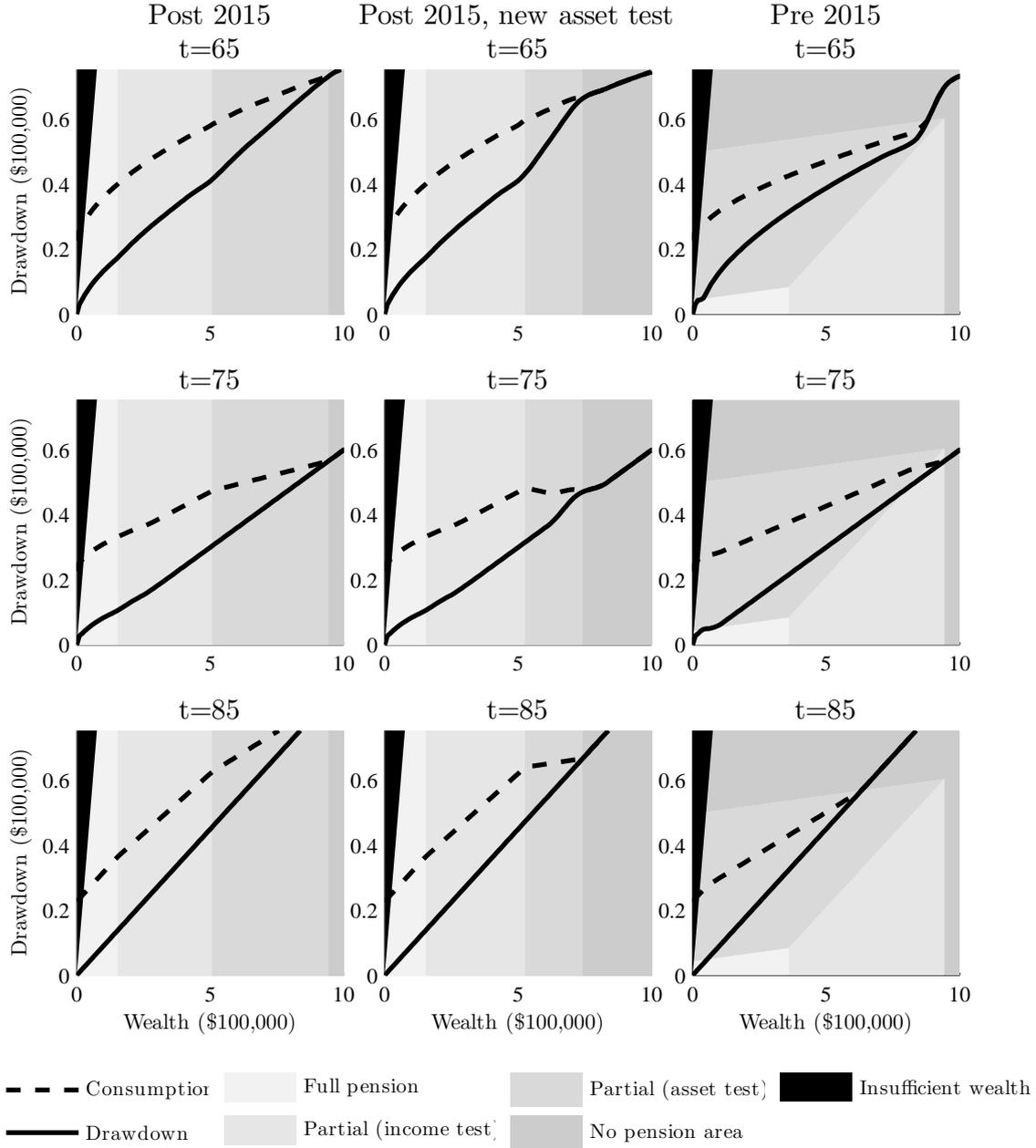}
  \captionof{figure}{Optimal drawdown and consumption for non-homeowner single households under the three different policy scenarios.}
  \label{fig:OptDD_S_W}
\end{figure}
 
Age Pension only contributes to the consumption, rather than being a means of planning for optimizing utility or amount of Age Pension received. This is in contrast to the results in \cite{Andreasson2016} with the policy rules prior to 2015, which showed that drawdown was highly sensitive to the means-test and could be utilized in financial planning (right column in Figure \ref{fig:OptDD_C_W}). There is a marginal effect when the retiree goes from no Age Pension to receiving partial Age Pension, especially for the 2017 asset-test adjustment, shown as a tiny dent where the consumption and drawdown curve intersect (the threshold between no pension and partial pension due to asset-test). This implies that a retiree should consume slightly more when his wealth is close to this threshold in order to receive partial Age Pension, but the additional utility would be so small that it is negligible in planning. Another level where such an optimizing decision would be expected is when the income-test binds over the asset-test (the threshold between partial pension due to income-test and asset-test). It occurs around \$508,066 for single households without a family home (\$248,352 for homeowners) and \$574,242 for couple households (\$314,527 for homeowners), and can be seen as a slight change in the drawdown curve due to different taper rates for the partial Age Pension. No apparent effect is identified in the consumption however, hence this would provide no additional utility. Note that as the retiree ages, and the minimum withdrawal rate is higher than the unconstrained optimal consumption, Age Pension simply adds to the consumption rather than being included in desired consumption. This is in line with \cite{Bateman2007}, which finds that welfare decreases slightly when minimum withdrawal rules are enforced over unconstrained optimal withdrawals, especially for higher levels of risk aversion.

\begin{figure}[!h]
\centering
\centerline{\includegraphics{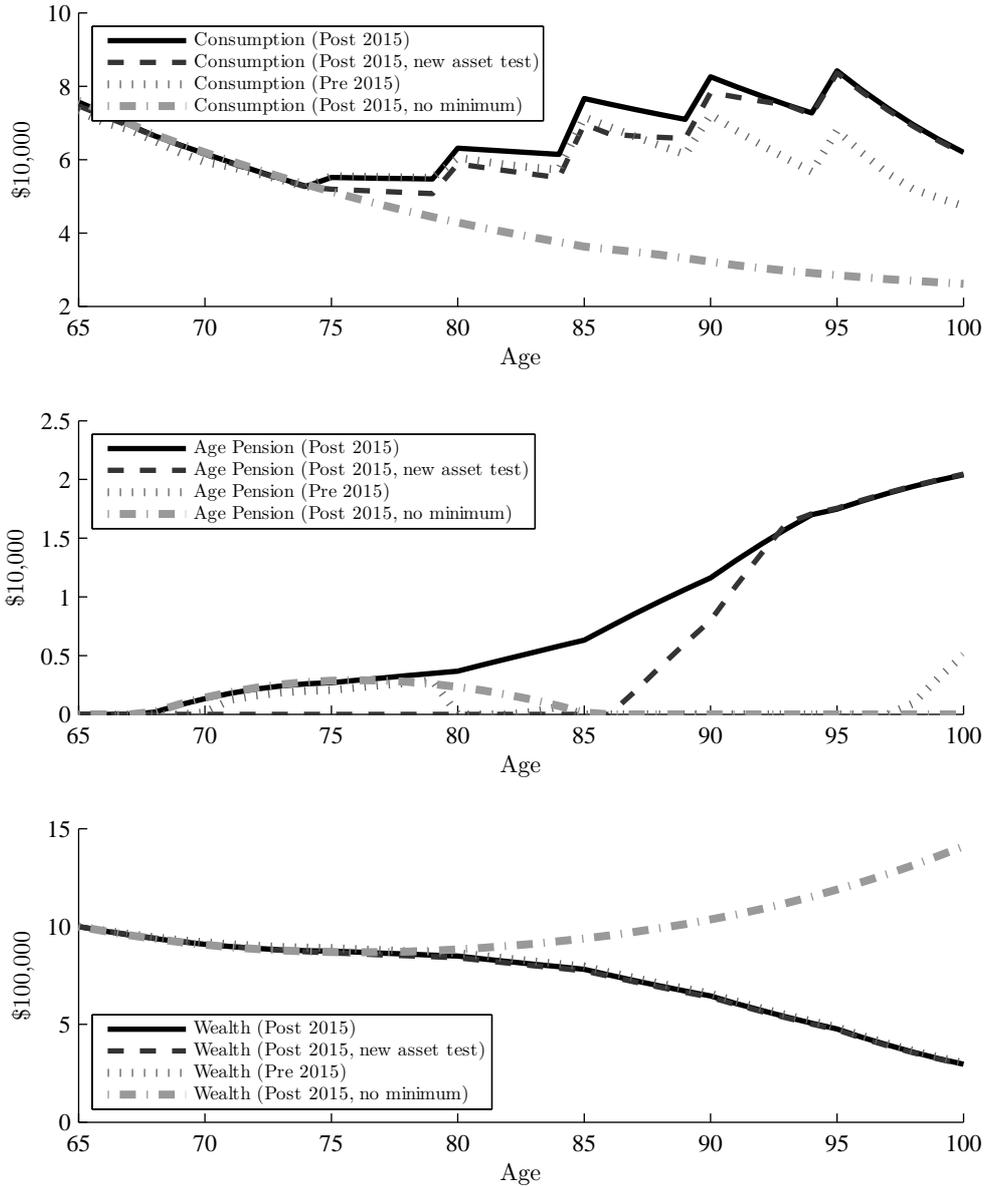}}
\caption{Comparison of consumption, Age Pension and wealth over a retiree's lifetime with the three different policy scenarios and with unconstrained (no minimum withdrawal) optimal consumption. The retiree starts with \$1m liquid wealth which grows with the expected return each year, and drawdown follows the optimal drawdown paths under each policy.}
\label{fig:LifetimeComparison}
\end{figure}

\begin{figure}[!h]
\centering
\centerline{\includegraphics{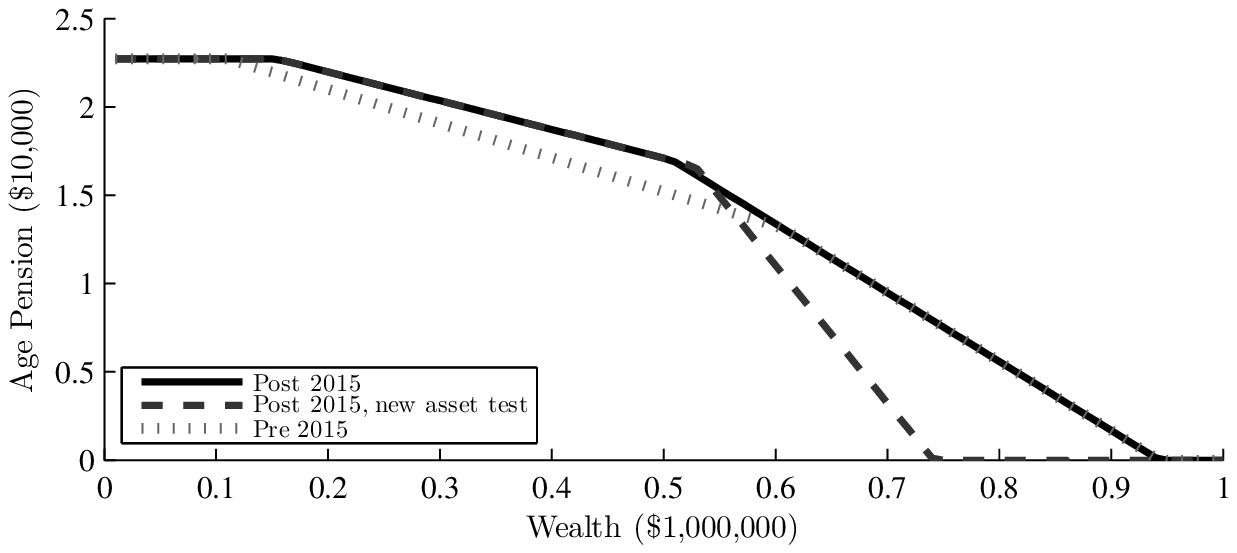}}
\caption{Comparison of the Age Pension function with the three policy scenarios. The retiree is a single household aged 65-74 and consumption is assumed to be the minimum withdrawal rate of 5\%.}
\label{fig:AgePensionFunction}
\end{figure}

An interesting outcome is when the consumption paths over a lifetime are compared with the new and old rules (Figure \ref{fig:LifetimeComparison}). Since the optimal drawdown rules are very similar before and after the change, and minimum withdrawal rates quickly binds, the consumption in turn follows the same pattern. However, since the income-test is now based on deemed income, more Age Pension is received in relation to wealth and drawdown assuming the deeming rates stay constant. This is especially true at older ages. Figure \ref{fig:AgePensionFunction} clearly shows the difference in Age Pension payments, where the new rules lead to more partial Age Pension (but less for wealthier households in the 2017 asset-test adjustment). As the withdrawal rate increases the difference in partial Age Pension increases as well. One of the reasons for changing the policy was for the government generate savings, but the deeming rules will not have the desired outcome on allocated pension accounts unless the deeming rates increase. Only when the minimum withdrawals are removed (or at least decreased), which in turn could lead to lower withdrawals for given wealth levels, could current rates lead to Age Pension payments being less under the new policy\footnote{It should be noted that the findings are for the account-based pension only, as other products which do not enforce the minimum withdrawal rates could incur additional savings for the government under the new rules.}. The effect can clearly be seen in Figure \ref{fig:LifetimeComparison}. Under the old rules, the relatively high drawdown (income) for the retiree would most often lead to no Age Pension due to the income-test, while under the new rules the retiree would receive a significant amount of Age Pension over his lifespan. Wealth paths throughout retirement, however, are almost identical - the difference between the new and old policy is solely in consumption from additional Age Pension.

\subsection{Optimal risky asset allocation}

\begin{figure}[!h]
\centering
\begin{minipage}{.5\textwidth}
  \centering
  \includegraphics[width=.9\linewidth]{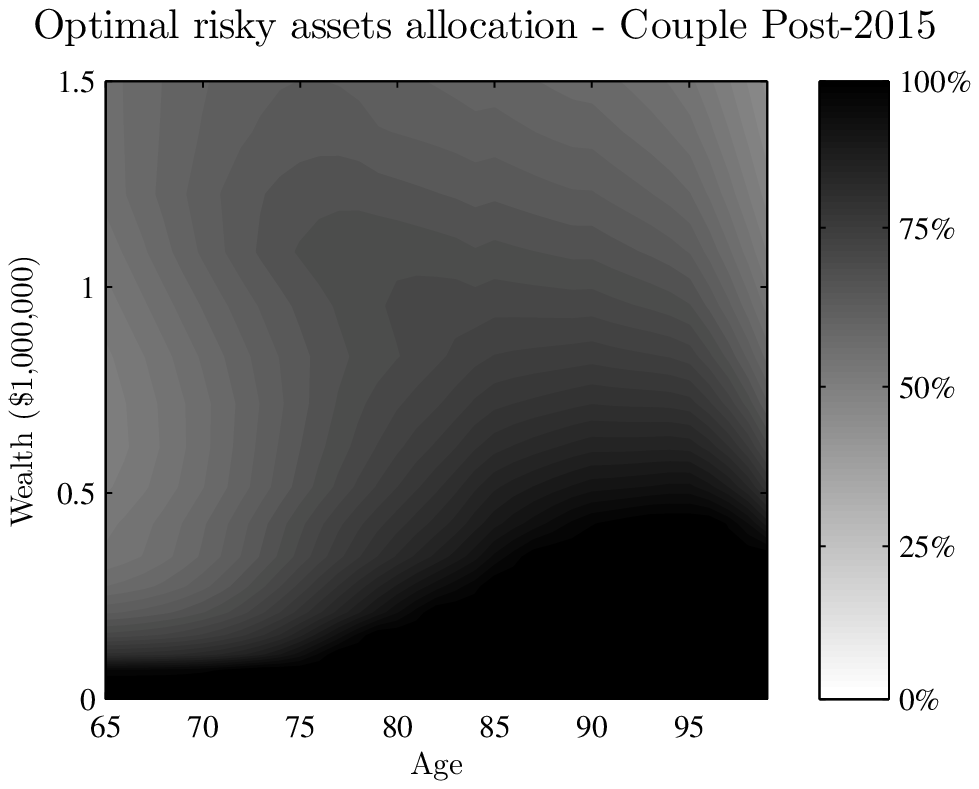}
  \captionof{figure}{Optimal risky allocation for non-homeowner couple household with post 2015 policy.}
  \label{fig:OptR_C_Post_W}
\end{minipage}%
\begin{minipage}{.5\textwidth}
  \centering
  \includegraphics[width=.9\linewidth]{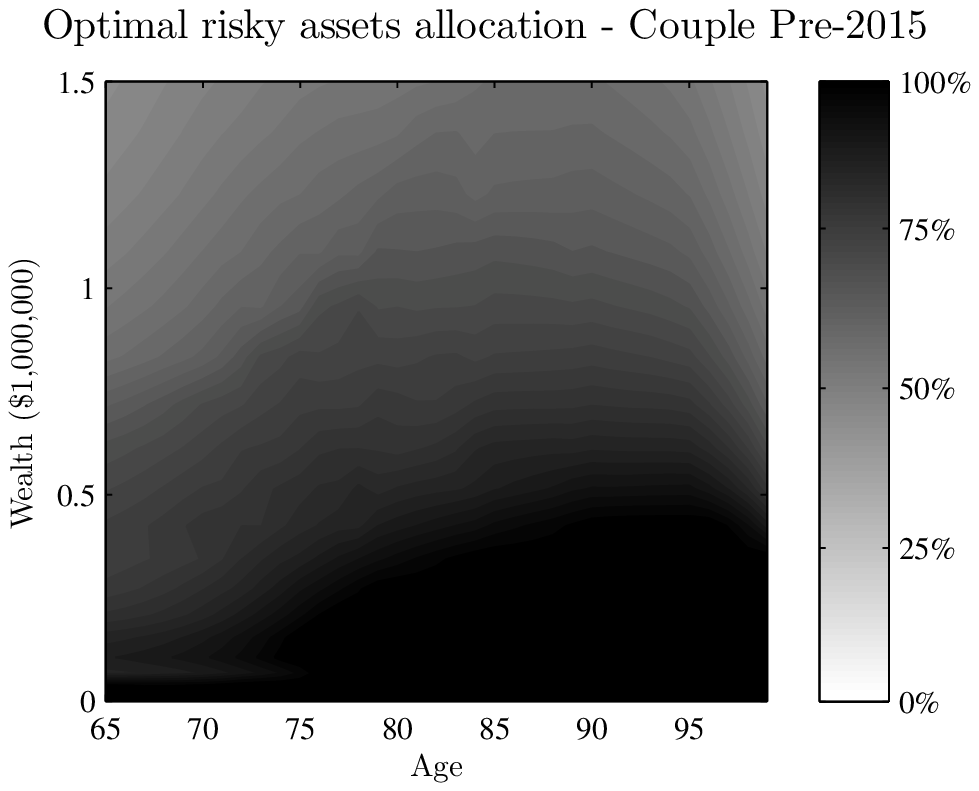}
  \captionof{figure}{Optimal risky allocation for non-homeowner couple household with pre 2015 policy.}
  \label{fig:OptR_C_Pre_W}
\end{minipage}
\end{figure}

\begin{figure}[!h]
\centering
\begin{minipage}{.5\textwidth}
  \centering
  \includegraphics[width=.9\linewidth]{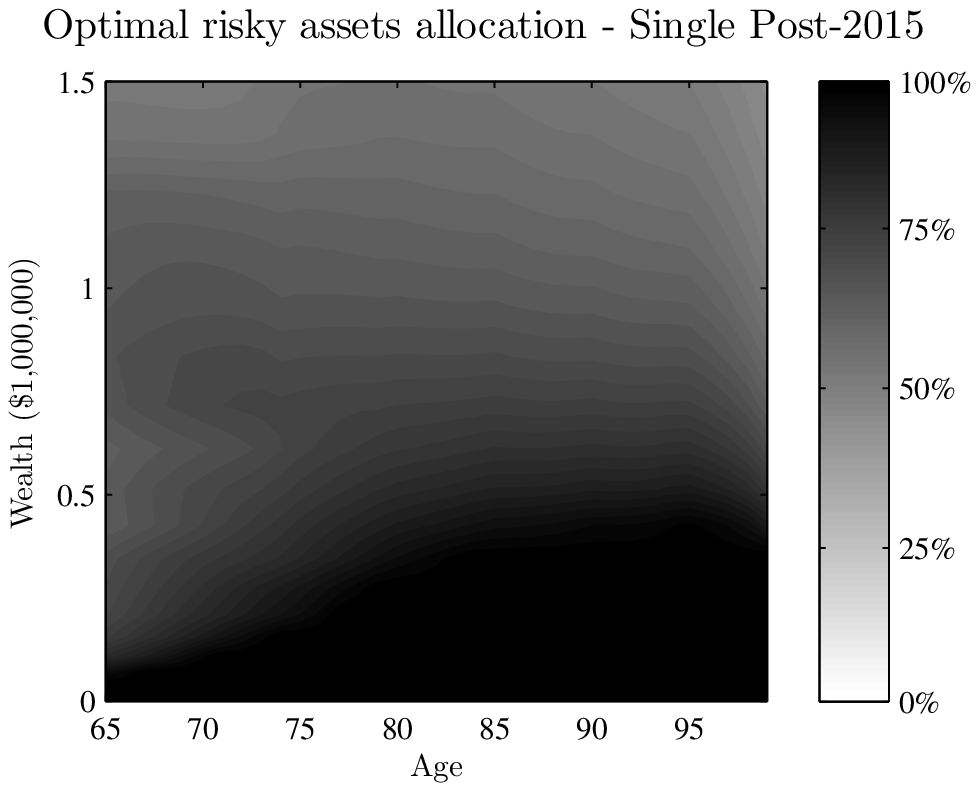}
  \captionof{figure}{Optimal risky allocation for non-homeowner single household with post 2015 policy.}
  \label{fig:OptR_S_Post_W}
\end{minipage}%
\begin{minipage}{.5\textwidth}
  \centering
  \includegraphics[width=.9\linewidth]{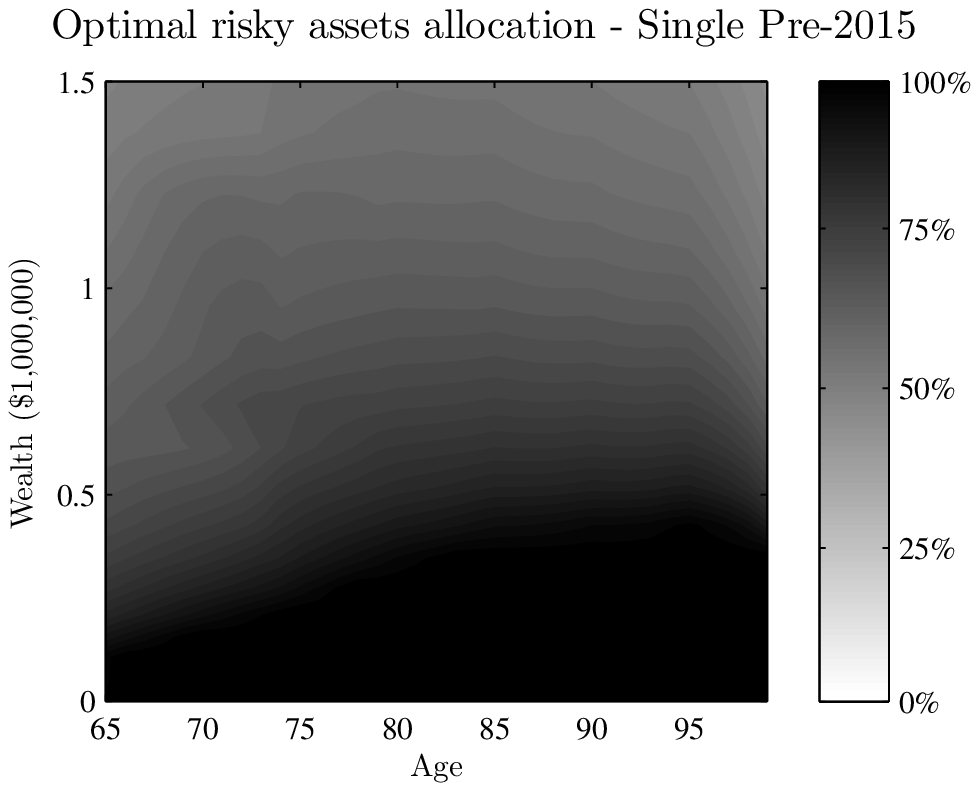}
  \captionof{figure}{Optimal risky allocation for non-homeowner single household with pre 2015 policy.}
  \label{fig:OptR_S_Pre_W}
\end{minipage}
\end{figure}

The exposure to risky assets in the portfolio is highly dependent on wealth and age, and even more so compared to the old rules. This is expected since the means-test is now based on wealth in both the asset and the income-test, which means investment returns will have a larger impact on expected utility. The risky allocation displays similar characteristics as the older rules and can be explained with the expected marginal utility conditional on wealth. When marginal utility increases with wealth, the risky allocation will always suggest 100\% risky assets. This is the case for the black bottom area (Figure \ref{fig:OptR_C_Post_W}-\ref{fig:OptR_S_Pre_W}), where the upper bound to the left indicates the maximum marginal utility from consumption, and the upper bound to the right is the maximum marginal utility from bequest. If utility from consumption is considered individually, then lower levels of wealth will have higher marginal utility. If marginal utility from bequest is instead isolated, the same effect will occur albeit at a higher level than for consumption ($\sim\$$450,000). It is, therefore, optimal up to these levels to allocate 100\% to risky assets, as the reward is larger than the risk.

The marginal utility is also affected by the means-test, as a result of the `buffer' effect. This buffer occurs when the decreasing wealth that stems from an investment loss is partially offset via increased Age Pension and can be seen as the comparatively darker area around the upper white line (indicating where partial pension becomes no pension) in Figure \ref{fig:OptR_C_Post_W} or \ref{fig:OptR_S_Post_W}. The buffer effect is, therefore, strongest for a retiree who has no Age Pension but is close to receiving partial Age Pension. An investment loss, in this instance, would be offset by partial Age Pension, whereas an investment profit would not cause the retiree to miss out on Age Pension that he would otherwise receive. The taper rate is steeper for the asset-test than the income-test (especially for 2017), hence marginal utility is lower when the asset-test is binding. For very low levels of wealth, the buffer effect is the opposite; investment losses can never lead to more than full Age Pension, and investment profits will decrease the amount of partial Age Pension received, which will result in lower marginal utility. 

Another interesting effect occurs as the minimum withdrawal rates cross above unconstrained optimal drawdown. When the retiree is forced to withdraw more from his account than is optimal to consume, the marginal utility drops significantly. This occurs approximately at age 75 for both single and couple households, though slightly later for less wealthy households. The marginal utility received from consumption is essentially zero after this age, thus the utility consists of an increasingly larger proportion bequest as the retiree ages (and mortality risk increases). This switch occurs where the bottom black area starts to increase towards the right bound, as it moves from utility from consumption to utility from bequest (this is more apparent for couple households in Figure \ref{fig:OptR_C_Post_W}). These characteristics are very similar to the surface generated by the old rules when minimum withdrawals are enforced. In fact, once the minimum withdrawal rates exceed the optimal drawdown, they become nearly identical. The difference is therefore only for the initial years of retirement, ages 65-80, because of how the income-test is constructed. In regards to the 2017 asset test changes, the buffer feature are slightly stronger, but the characteristics are similar to the 2015 rules, hence are not shown in a graph. With the old rules, the income-test is binding most of the time, whereas with the new rules it binds for only roughly one-third of the partial Age Pension --- and even less than that for homeowners.

\subsection{Optimal housing allocation}
\begin{figure}[!h]
\centering
\centerline{\includegraphics{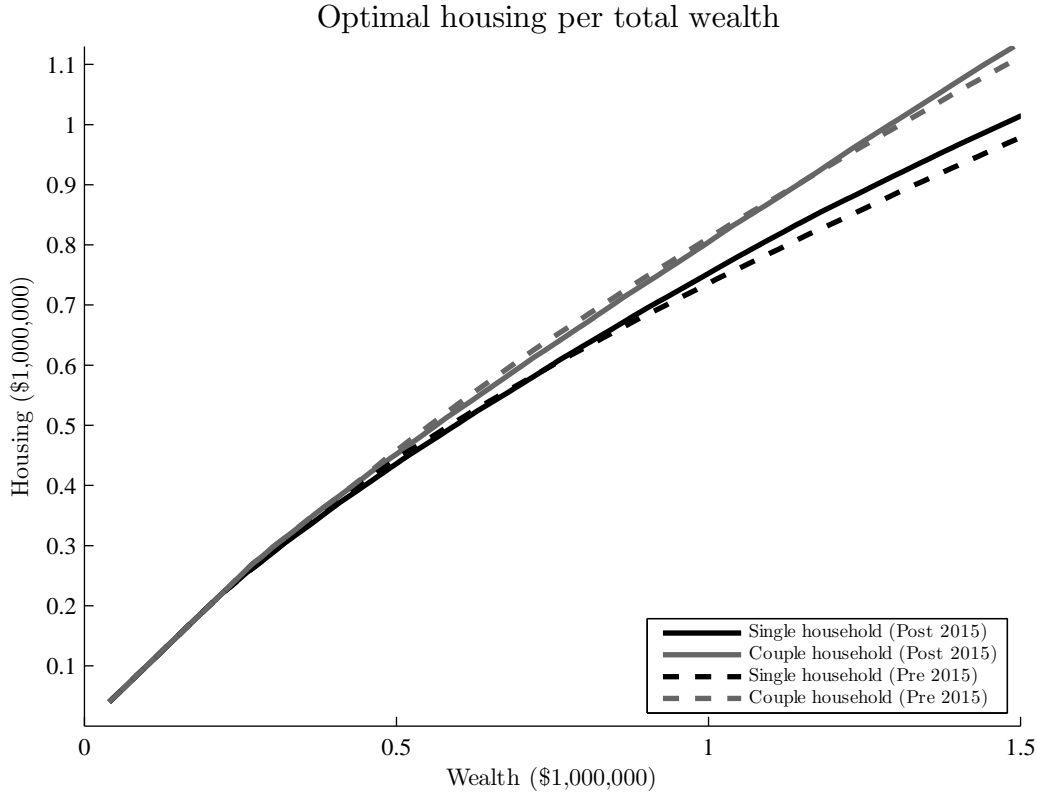}}
\caption{Optimal housing allocation given by total wealth $\mathsf{W}$ for single and couple households under the pre 2015 and post 2015 policy.}
\label{fig:Housing}
\end{figure}

The decision variable for the allocation of assets into a family home is expected to  change slightly due to the increased focus on assets in the means-test. The decision made at the time of retirement shows that under the new policy rules it is optimal to invest slightly less than under the old rules, up to a total wealth level of approximately \$735,000 for single households and \$1,155,000 for couple households (see Figure \ref{fig:Housing}). This would leave approximately \$144,000 and \$247,000 respectively as liquid wealth. Households with total wealth above this level, meanwhile, are recommended to invest slightly more. These allocation decisions leave liquid wealth just below the thresholds for receiving full Age Pension, and the difference in the housing curves can be explained by the income-test changes. For a given wealth, the new rules provide the retiree with more partial pension than with the old rules. Early in retirement, the optimal consumption is high which causes the income-test to bind under the old rules. The new rules alternatively have a deemed income for the income-test that is much lower than before, which ultimately results in more partial Age Pension. The effect decreases 10-15 years into retirement, but when the minimum withdrawal rates exceed unconstrained optimal drawdown at older age ranges, the same occurs again. This can be seen by comparing plots in column 1 or 2 with column 3 in Figure \ref{fig:OptDD_C_W} or Figure \ref{fig:OptDD_S_W}. Since a certain level of liquid wealth under the new rules will lead to higher expected utility, it is optimal to allocate slightly more in housing compared with the old rules (as long as the liquid wealth is not very low) to benefit from receiving partial Age Pension. The effects of the 2017 asset test changes are very similar and cannot be distinguished visually from the 2015 policy, hence have been left out in Figure \ref{fig:Housing}.

\section{Conclusions}
In this paper, we adapt the stochastic retirement utility model from \cite{Andreasson2016} to implement the Age Pension policy changes from 2015, which affect all Allocated Pension accounts opened after January 1\textsuperscript{st} 2015. These changes affect the treatment of income for the Age Pension income-test, and lead to different optimal decisions for consumption, investments, and housing. We also evaluate the new policy rules with the current Age Pension asset-test, as well as the planned asset-test adjustments for 2017.

We find that optimal consumption only applies early in retirement, as minimum withdrawal rates exceed unconstrained optimal drawdown rates between ages 75-85, depending on wealth level. Only before this point, it is possible to plan withdrawals in order maximize utility, but these possibilities are almost nonexistent under the new policy rules compared with previous. Optimal drawdown equals minimum withdrawal after age 85 (as it becomes a binding lower constraint for withdrawal), thus the new and old policy rules are identical after this age. That said, since the income-test tends to bind for the old rules while the asset-test dominates for the new rules, the retiree will now tend to receive more partial pension under the optimal withdrawal rules. Even with the steeper taper rate that will be introduced in January 2017, the retiree will receive a more generous Age Pension compared with the old policy.

Since income (which was considered drawdown from the allocated pension account) is now replaced by deemed income, the assets are means-tested twice, which means risky asset allocation becomes more sensitive. The changes in optimal risky asset allocation over time and wealth are similar to the old rules, but the changes are slightly more aggressive and depend on marginal utility from consumption and bequest, as well as the level of buffering against investment losses the Age Pension provides. This effect dies off as the minimum withdrawal rates bind, and the bequest motive becomes more important.

Providing that the retiree's remaining liquid wealth is close to (or higher) than the threshold between full and partial Age Pension at the time of retirement, it is optimal to invest slightly more in housing than before. This will allow the retiree to receive more partial Age Pension, and to increase his expected utility in the long term. If the retiree instead has lower total wealth than the threshold, he is alternatively recommended to invest marginally less than before.

One surprising finding is that a retiree with an income stream where minimum withdrawal rules are enforced will receive more Age Pension over the course of their lifetime with the new policy rules. Due to the minimum withdrawal requirement, the drawdown tends to be higher than what is optimal for most ages, which under the old rules would result in no or low partial Age Pension. The new rules combined with the current historically low deeming rates will generate significant Age Pension payments from the same drawdown and wealth levels. This, in turn, affects both the decision for allocation in housing as well as risky investments. The government's goal of reducing incentives for maximizing Age Pension payments and focusing on maximizing total disposable income is however met - the new policy is not as sensitive to optimal withdrawal decisions in order to maximize Age Pension payments as the old policy was.

\section*{Acknowledgment}
This research was supported by the CSIRO-Monash
Superannuation Research Cluster, a collaboration among CSIRO, Monash University, Griffith University, the University of Western Australia, the University of Warwick, and stakeholders of the retirement system in the interest of better outcomes for all. Pavel Shevchenko acknowledges the support of Australian Research Council's Discovery Projects funding scheme (project number DP160103489).

\bibliographystyle{te}

{\footnotesize
\bibliography{bibliography}}

\end{document}